\documentstyle[preprint,prb,aps,epsf]{revtex}
\begin{document}

\title{
Effects of ${\mathbf D}$-strain, ${\mathbf g}$-strain,
and dipolar interactions on EPR linewidths of the molecular magnets
Fe$_{\mathbf 8}$ and Mn$_{\mathbf 12}$ }
%\title{{\bf ${\mathbf D}$-strain effect on EPR Linewidths of 
%the Nanomagnets
%${\mathbf Fe_8}$ and ${\mathbf Mn_{12}}$}}
% \draft command makes pacs numbers print
\draft 
% repeat the \author\address pair as needed
\author{Kyungwha ${\mathrm Park^{1,\ast}}$ 
\and M.~A.\ ${\mathrm Novotny^{1,\dag}}$ 
\and N.~S.\ ${\mathrm Dalal^2}$ \and S.\ ${\mathrm Hill^{3,\ddag}}$, 
and P.~A.\ ${\mathrm Rikvold^{1,4}}$}
\address{${\mathrm ^{1}}$School of Computational Science and 
Information Technology, Florida State University, Tallahassee, FL 32306 \\
${\mathrm ^{2}}$Department of Chemistry, Florida State University,
Tallahassee, FL 32306 \\
${\mathrm ^{3}}$Department of Physics, Montana State University,
Bozeman, MT 59717 \\
${\mathrm ^{4}}$Center for Materials Research and 
Technology and Department of Physics, 
Florida State University, Tallahassee, FL 32306}
%\author{Kyungwha ${\mathrm Park^{1\ast}}$ 
%\and M.~A.\ ${\mathrm Novotny^2}$ 
%\and N.~S.\ ${\mathrm Dalal^3}$ \and S.\ ${\mathrm Hill^{4}}$, 
%and P.~A.\ ${\mathrm Rikvold^{1,5}}$}
%\address{${\mathrm ^{1}}$School of Computational Science and 
%Information Technology, Florida State University, Tallahassee, FL 32306 \\
%${\mathrm ^{2}}$Department of Physics and Astronomy, Mississippi 
%State University, Mississippi State, MS 39762 \\
%${\mathrm ^{3}}$Department of Chemistry, Florida State University,
%Tallahassee, FL 32306 \\
%${\mathrm ^{4}}$Department of Physics, 
%University of Florida, Gainesville, FL 32611 \\
%${\mathrm ^{5}}$Center for Materials Research and 
%Technology and Department of Physics, 
%Florida State University, Tallahassee, FL 32306}
\date{\today}
\maketitle

\begin{abstract}
Electron paramagnetic resonance (EPR) measurements on single crystals
of the molecular magnets ${\mathrm Fe_8}$ and ${\mathrm Mn_{12}}$ 
reveal complex
nonlinearities in the linewidths as functions of energy eigenstate,
frequency, and temperature. Using a density-matrix equation with
distributions of the uniaxial anisotropy parameter $D$,
the Land\'{e} $g$ factor, and dipolar interactions, we obtain
linewidths in good agreement with experiments.
Our study shows that the distribution in $D$ is common to
the examined molecular magnets Fe$_8$ and Mn$_{12}$ regardless
of the qualities of the samples. This 
%The standard deviation of $D$ is estimated as $0.01D$ ($0.02D$) 
%for ${\mathrm Fe_8}$ (${\mathrm Mn_{12}}$), which 
could provide the basis for a proposed tunneling
mechanism due to lattice defects.
The distribution in $g$ is also quite significant for Mn$_{12}$.

%supports a recently proposed tunneling mechanism.
\end{abstract}
\pacs{PACS numbers: 75.45.+j, 75.50.Xx, 76.30.-v}

%\begin{multicols}{2}

\section{Introduction}

Molecular magnets such as ${\mathrm Mn_{12}}$
\cite{LIS80} and ${\mathrm Fe_8}$ \cite{WIEG84} have recently 
drawn vigorous attention 
because of the macroscopic quantum tunneling (MQT) of their magnetizations 
at low temperatures \cite{VILL94,CHUD98} and their possible applications
to quantum computing.\cite{LEUE00-3}  
These materials consist of many identical clusters with the same 
magnetic properties and characteristic energies.  
Each cluster is made up of
many different species of ions and atoms, with a
total spin angular momentum in the ground state of $S$$=$$10$.  
The clusters have strong crystal-field anisotropy and, thus,
a well-defined easy axis, 
and the magnetic interaction between different clusters is weak.
Many competing models have been proposed to 
explain MQT in molecular magnets: higher-order transverse 
anisotropy,\cite{HART96} thermally-assisted quantum 
tunneling,\cite{FRIE96,LEUE00-1} the Landau-Zener 
effect,\cite{LEUE00-2,MIYA97,DOBR97}
and dipolar interactions with dynamic hyperfine fields.\cite{PROK98} 
So far, the MQT in these materials has not been completely understood,
and new experimental and theoretical approaches are needed.

In this paper, we present both previously unpublished 
experimental data and a theoretical
model for the electron paramagnetic resonance (EPR) linewidths of 
single crystals of the molecular magnets Fe$_8$ and Mn$_{12}$ when the field
is applied along the easy axis. To understand the measured linewidths,
we use density-matrix equations, 
assuming that the uniaxial anisotropy parameter $D$ and 
the Land{\'e} $g$ factor are randomly distributed around their 
mean values (``$D$-strain'' and ``$g$-strain'' effects \cite{PILB})
due to crystalline defects or impurities in samples.
By adjusting the standard deviations of $D$ and $g$, and the
strength of the dipolar interactions, 
these equations are used to obtain theoretical linewidths in 
excellent quantitative agreement with our experimental 
data. Our theoretical study also shows that 
for the examined Fe$_8$ sample the distribution in $D$ and 
the dipolar interactions are crucial to understand the linewidths, 
while for the Mn$_{12}$ sample the distributions in $D$ and $g$ 
are more important than the dipolar interactions. 
These results imply that the distribution in $D$ seems to be 
generic in the molecular magnets Fe$_8$ and
Mn$_{12}$, although the standard deviation of $D$ itself varies
between samples. This point assures that inherent defects or
impurities in samples can be checked through the distribution in $D$
by observing the linewidths of EPR spectra, and forms a basis 
of a recently proposed dislocation-induced tunneling mechanism.
\cite{CHUD01}

The experimental details are presented in Sec. II. Section III
describes some typical EPR spectra and observed features, while
Sec. IV presents our theoretical model. Section V discusses
the theoretical results in comparison with the experimental data.
Section VI describes our conclusions.

\section{Experimental Details}

All the measurements were made on single crystals. Mn$_{12}$ was
synthesized according to the procedure described by Lis.\cite{LIS80}
Fe$_8$ was synthesized using Wieghart's method.\cite{WIEG84} Single
crystals of both materials were grown via slow evaporation.
Typical crystal dimensions used were 1$\times$1$\times$0.5mm$^3$.
EPR measurements were made for a magnetic field along or close to the easy axis
using the multifrequency-resonator-based spectrometer described 
earlier.\cite{HILL98,MACC01,PERE98,HILL99}
The (transverse) excitation frequency was varied between 45 and 190 GHz. 
The temperature range was from 1.8 K to 40 K, with a precision of
0.05 K. The linewidths were measured as functions of
Zeeman field, EPR frequency, and temperature, 
for crystals of different quality, over a wide enough 
range to probe energy-level broadening due to lattice defects and
inter-cluster dipolar interactions. 
 
\section{Experimental Results}

\begin{figure}
\begin{center}
%\leavevmode
\epsfysize=8.cm
\epsfysize=6.cm
\epsfbox{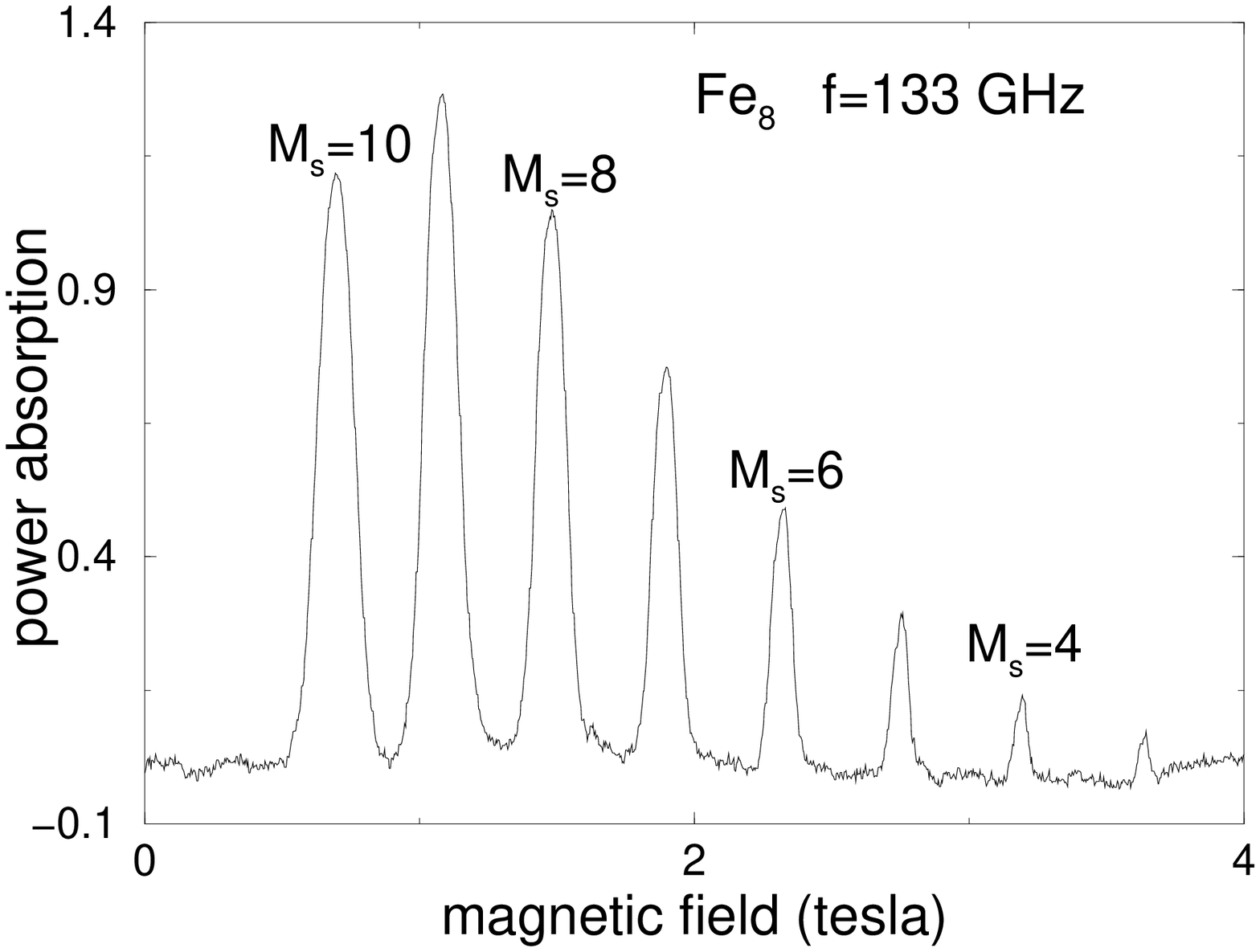}
\caption{Normalized power absorption (in arbitrary units) 
for ${\mathrm Fe_8}$ as a function of magnetic field 
at $T$$=$$10~{\mathrm K}$ with the field almost along 
the easy axis ($\theta=9^{\circ}$) \protect\cite{MACC01}. 
Every other resonance peak is marked by the energy level
from which the spin system is excited. For example,
the leftmost peak is for the transition
$M_s$$=$10$\rightarrow$9.}
\label{pws_Fe8}
\end{center}
\end{figure}

\begin{figure}
\begin{center}
%\leavevmode
\epsfxsize=8.cm
\epsfysize=6.cm
\epsfbox{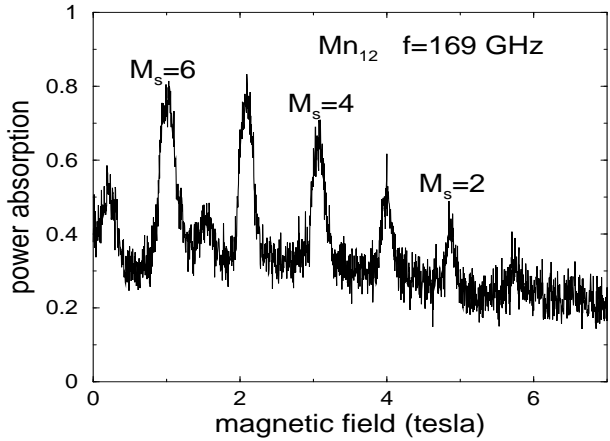}
\caption{Normalized power absorption (in arbitrary units) 
for Mn$_{12}$ as a function of magnetic field 
at $T$$=$$25~{\mathrm K}$ with the field along
the easy axis. Every other resonance peak is marked by 
the energy level from which the spin system is excited.
The leftmost full resonance peak around
1 T is for the transition $M_s$$=$6$\rightarrow$5.}
\label{pws_Mn12}
\end{center}
\end{figure}

Figure~\ref{pws_Fe8} shows a typical EPR spectrum of Fe$_8$ for a field
approximately along the easy axis at a resonance frequency of 133 GHz and at
$T$$=$10 K.\cite{MACC01} Figure~\ref{pws_Mn12} shows a 
typical EPR spectrum of Mn$_{12}$ for a field along the 
easy axis at a resonance frequency of 169 GHz and $T$$=$25 K.
The following interesting features are observed:
the EPR linewidths are on the order of several hundred to a thousand
gauss for ${\mathrm Fe_8}$ (one to a few thousand gauss for
${\mathrm Mn_{12}}$) [Figs.~\ref{pws_Fe8},~\ref{pws_Mn12},
\ref{FWHM}(a),(b),(c), and (d)]; 
the linewidths increase {\it non-linearly} as a 
function of the absolute value of the energy eigenstate $M_s$
[Figs.~\ref{pws_Fe8},~\ref{pws_Mn12},~\ref{FWHM}(a),(b),(c), and (d)]; 
the minimum linewidths are observed for transitions involving the 
$M_s$$=$$0$ eigenstate,
with a weak asymmetry in the widths about $M_s$$=$$0$
[Figs.~\ref{FWHM}(a) and~(b)]; and resonances involving
the same energy eigenstates become narrower upon increasing the
measurement frequency or field (compare Fig.~\ref{FWHM}(a) 
with \ref{FWHM}(c)). 

\section{Model}

For clarity, the model for Fe$_8$ is described separately from
that for Mn$_{12}$.

\subsection{Fe$_8$}

To obtain the theoretical linewidths of the power absorption 
for ${\mathrm Fe_8}$ clusters, 
%To obtain the theoretical EPR linewidths for ${\mathrm Fe_8}$ clusters,
we consider a single-spin system with $S$$=$$10$ in a weak oscillating
transverse magnetic field. We choose the easy axis to be the $z$-axis.  
Since the ${\mathrm Fe_8}$ clusters have an approximate $D_2$ symmetry,
the lowest-order ground-state single-spin Hamiltonian is \cite{AABB}
\begin{eqnarray}
{\cal H}_{0} &=& -D S_z^2 - E(S_x^2 - S_y^2) 
- g \mu_B H_z S_z ~, \label{eq:ham}
\end{eqnarray}
where $D$$=$$0.288k_B$ ($k_B$ is Boltzmann constant) and the transverse 
crystal-field anisotropy parameter, $E$, is $0.043k_B$.\cite{MACC01}
Here $S_{\alpha}$ is the $\alpha$-th component of 
the spin angular momentum operator, $g$ is 
the Land\'{e} $g$-factor which is close to 2, $\mu_B$ is the Bohr magneton,
and $H_z$ is the longitudinal static applied magnetic field.
In zero field, ignoring the small transverse anisotropy terms, 
the eigenstates 
$(1/ \sqrt{2})(|M_s \rangle \pm |$$-$$M_s$$ \rangle)$
are degenerate for each $M_s$, and the lowest energy states 
are $(1/ \sqrt{2})(|10 \rangle \pm |$$-$$10$$ \rangle)$.  
These degeneracies are lifted by applying the field, $H_z$.
When $H_z \approx D(m+m^{\prime})/g \mu_B$, where
$m$ and $m^{\prime}$ are eigenvalues of $S_z$
in units of $\hbar$ ($m,m^{\prime}$$=$$-10, \cdots , 10$),
there exists a pair of quasi-degenerate energy levels, $m$ and $m^{\prime}$.
When $H_z$ is large enough, the eigenvalue of $S_z$, $M_s$, is a good
quantum number since the transverse terms are much
smaller than the longitudinal terms. 
In such magnetic fields, the transverse terms, $E(S_x^2 - S_y^2)$,
cause quantum tunneling between the two quasi-degenerate energy levels.
Next we introduce an interaction $V(t)$ between the spin system and an
oscillating transverse magnetic field $H_x$ with angular frequency 
$\omega$$\equiv$$ 2 \pi f$:
\begin{eqnarray}
V(t) &=& \frac{V_0}{2}(e^{i \omega t}+e^{-i \omega t})~, 
~~~~V_0 ~\propto~  H_x S_x ~. 
\label{eq:inter}
\end{eqnarray}
We treat $V(t)$ as a small perturbation to ${\cal H}_0$.

Besides the interaction $V(t)$, the spin system interacts 
with the environment, such as the thermal fluctuations of the clusters.  
Through this spin-phonon coupling, 
the spin system can relax to re-establish thermal equilibrium.  
Since the times of interest are much longer than the correlation 
time of the phonon heat bath, the heat bath is considered always to be 
in thermal equilibrium. Thus we can integrate out all degrees of freedom
of the bath to obtain an equation of motion for the 
density matrix $\rho(t)$ of the spin system:\cite{BLUM}
\begin{eqnarray}
\frac{d \rho(t)_{m^{\prime},m}}{dt} &=& \frac{i}{\hbar} 
[\rho(t),{\cal H}_{0}+V(t)]_{m^{\prime},m} \nonumber \\
& & + \delta_{m^{\prime},m} \sum_{n\neq m} \rho_{n,n} W_{mn} 
- \gamma_{m^{\prime},m} \rho_{m^{\prime},m}~,
\label{eq:GME} \\
\gamma_{m^{\prime},m} &=& \frac{W_m + W_{m^{\prime}}}{2}~, ~~~~
W_m ~=~ \sum_{k \neq m} W_{km}~,
\end{eqnarray}
where the subscripts represent eigenstates of 
the longitudinal part of ${\cal H}_0$, $\rho(t)_{m^{\prime} m}$
$=$$\langle m^{\prime}| \rho(t) | m \rangle$, and 
$W_{km}$ is the transition rate 
from the $m$-th to the $k$-th eigenstate, which 
is determined by the spin-phonon coupling.\cite{LEUE00-2}

We consider the case that the frequency $\omega$ of the oscillating
field is fixed while $H_z$ is varied to induce a resonance.  
With the selection rule $\Delta M_s$$=$$\pm 1$, 
the power absorbed between the $M_s$ 
and the $M_s-1$ level is
\begin{eqnarray}
\frac{d{\cal E}}{dt} 
&=&{\cal E}_{M_s} [\dot{\rho}_{M_s,M_s}]_{\rm rad} 
+  {\cal E}_{M_s-1} [\dot{\rho}_{M_s-1,M_s-1}]_{\rm rad}~,
\end{eqnarray}
where ${\cal E}_{M_s}$ is the energy of the $M_s$ level, and 
$[\dot{\rho}_{M_s,M_s}]_{\rm rad}$ is the change with time
of the population in the level $M_s$ caused by the first term in 
Eq.~(\ref{eq:GME}). We solve Eq.~(\ref{eq:GME}) 
for $[\dot{\rho}_{M_s,M_s}]_{\rm rad}$ up 
to first order in $V_0$ near resonance 
[$\omega \approx ({\cal E}_{M_s} - {\cal E}_{M_s-1})/\hbar$].  
In this limit, among the many off-diagonal density-matrix elements,
only $\rho(t)_{M_s-1,M_s}$ [$=\rho(t)_{M_s,M_s-1}^{\ast}$] contributes to
$[\dot{\rho}_{M_s,M_s}]_{\rm rad}$. The matrix element $\rho(t)_{M_s-1,M_s}$
is obtained by solving Eq.~(\ref{eq:GME}). Consequently, only 
two neighboring energy levels are involved for each resonance peak.   
An increase of the population in one of the two energy levels results
in a decrease of the population in the other.
%$[ \dot{\rho}_{M_s,M_s} ]_{\rm rad}$$=$$ 
%- [ \dot{\rho}_{M_s+1,M_s+1} ]_{\rm rad}$.  
So the power absorption becomes \cite{BLUM} 
\begin{eqnarray}
\! \frac{d{\cal E}}{dt} \! &=& \! 
\frac{({\cal E}_{M_s-1} \! - \! {\cal E}_{M_s})}{\hbar^2} {\tilde V}^2 
\Delta(H_z) (\rho_{M_s,M_s} \! - \! \rho_{M_s-1,M_s-1})~, \nonumber \\
 \Delta(H_z)&\equiv& \frac{\hbar^2 \gamma_{M_s-1,M_s}}
{ (g \mu_B)^2 ( H_z - H_{\rm res})^2 + ( \hbar \gamma_{M_s-1,M_s} )^2 } ~,
\nonumber \\ 
\! \!  {\tilde V} &\! \equiv \! &|\langle M_s|V_0|M_s-1 \rangle|,
~H_{\rm res} \equiv \frac{\hbar \omega - D(2 M_s - 1)}{g \mu_B} \;,
\label{eq:res_field} 
\end{eqnarray}
where $\Delta(H_z)$ is a Lorentzian lineshape function, 
$H_{\rm res}$ is the resonant field, 
and $\hbar \gamma_{M_s-1,M_s}/g \mu_B$ gives a linewidth
due to the finite lifetime of any excited state.
As $M_s$ decreases, the resonant field increases. 
The linewidth determined by $\gamma_{Ms-1,Ms}$
is on the order of several to several tens of gauss at temperatures
below several tens of kelvin, and it increases as $M_s$ decreases
because of the corresponding shorter lifetimes.
However, the experimentally observed linewidths {\em decrease\/} 
as $M_s$ decreases until the
$M_s$$=$$1 \rightarrow 0$ and $M_s$$=$$0 \rightarrow -1$ transitions
(hereafter denoted $M_s$$=$1 and $M_s$$=$0, respectively), 
and are on the order of several hundred 
to a thousand gauss in the same temperature range 
[Figs.~\ref{pws_Fe8} and~\ref{FWHM} (a),(b), and (c)].

To resolve these large discrepancies, we first assume that 
$D$ and $g$ are independent random variables
with Gaussian distributions centered at $0.288k_B$ and $2.00$, with
standard deviations $\sigma_D$ and $\sigma_g$, respectively.  
Such distributions can be caused by impurities or crystalline
defects in the macroscopic sample, which result in different
clusters seeing slightly different values of $D$ and $g$.
%(``$D$-strain'' and ``$g$-strain'' \cite{PILB}).
Then we calculate the average power absorption at a fixed frequency 
and $T$$=$$10~{\mathrm K}$ by averaging Eq.~(\ref{eq:res_field}) 
over the Gaussian distributions
using {\texttt{Mathematica}} \cite{MATH}, to obtain 
the linewidth as a function of $M_s$.  
%We vary $\sigma_D$ and $\sigma_g$ to compare with 
%the experimental data \cite{MACC01}.
%The lineshape depends on the magnitude of $\sigma_D$ and $\sigma_g$,
%compared to the natural linewidth determined by $\gamma_{M_s-1,M_s}$.  
%If $\sigma_D$ and $\sigma_g$ are much larger than (comparable to) 
%the natural linewidth, then the resulting lineshape is 
%Gaussian (Lorentzian). Otherwise, the absorption linewidths
%do not fit to either a Gaussian or a Lorentzian. 
Our numerical calculations show that 
the distribution in $D$ causes the linewidths to decrease linearly 
with decreasing absolute value of $2M_s$$-$$1$, with a
slight rounding close to the linewidth minimum ($M_s$$=$$1,0$),
where the $D$-strained linewidths approach the intrinsic
lifetime broadening.
%flatten out at $M_s$$=$$1,0$, and then rise again for $M_s < 0$. 
On the other hand,
the distribution in $g$ makes the linewidths increase with 
decreasing $M_s$ because the resonant field increases 
with decreasing $M_s$. These trends can be predicted 
from the expression for the resonant field $H_{\mathrm res}$ 
in Eq. (\ref{eq:res_field}) 
because the intrinsic linewidths are substantially smaller than the
measured linewidths (with the exception of the lowest absolute
$M_s$ transitions). However, the density-matrix 
equation (\ref{eq:GME}) is needed to understand the temperature dependence 
of the linewidths. 

%Next, we consider the effect of the inter-cluster 
%dipolar interactions. 
%These make each cluster see a different 
%net magnetic field, which results in a broadening of the linewidths.
%When the thermal energy becomes comparable to (or smaller than) 
%the energy of the oscillating field, some energy levels are not 
%populated, so the dipolar interactions become temperature dependent.
Next, we consider the effect of the inter-cluster dipolar interactions,
in which each cluster experiences a net magnetic field comprised of
the applied field and the dipolar fields from the surrounding clusters.
At high temperatures ($k_B T$$\gg$$\hbar \omega$) these dipolar fields
fluctuate randomly about zero, giving rise to a dipolar contribution
to the EPR linewidths. At low temperatures, the ground state
becomes preferentially populated, leading to nonzero dipolar fields
which are the same at the site of each cluster. This leads to 
%a line shift and, more importantly, 
a line narrowing upon increasing
$H_{\mathrm res}$ or the EPR frequency, or reducing the temperature.
We use the multi-spin Hamiltonian that
consists of ${\cal H}_0$ and $V(t)$ for each cluster,
and magnetic dipolar interactions between different clusters. 
We neglect the spin-phonon interaction because the spin-phonon 
relaxation time is very large compared to the spin-spin relaxation time.
%With the selection rule $\Delta M_s$$=$$\pm 1$, we truncate
%the original multi-spin Hamiltonian to leave 
%only terms which commute with the $z$ component of the total
%spin angular momentum $\sum_{j} S_{zj}$, where
%the sum runs over all clusters. 
We truncate the original multi-spin Hamiltonian to leave 
only terms which commute with the $z$ component of the total
spin angular momentum $\sum_{j} S_{zj}$, where
the sum runs over all clusters, because of the selection rule
$\Delta \tilde{M}_s$$=$$\pm 1$ where $\tilde{M}_s$ is an eigenvalue
of $\sum_{j} S_{zj}$.
Then the dipolar interaction
is modified to ${\cal H}^{{\mathrm dp}}=\sum_{k > j} A_{jk} 
( \vec{S}_j \cdot \vec{S}_{k} - 3  S_{zj} S_{zk} )$, 
where $A_{jk} \equiv [(g \mu_B)^2/ 2 r_{jk}^3] 
(3 \zeta_{jk}^2 - 1)$.  Here $r_{jk}$ is the distance between
clusters $j$ and $k$, and $\zeta_{jk}$ is the direction 
cosine of $r_{jk}$ relative to the $z$-axis. 
%Each cluster is allowed to have $S_x$ or $S_y$ 
%as far as the selection rule is satisfied. 
Assuming that 
$\sum_{j} V_j(t)$$\ll$${\cal H}^{{\mathrm dp}}$$\ll$$\sum_{j}{\cal H}_{0j}$,
we neglect $\sum_{j} V_j(t)$ in the energy of the multi-spin 
system and treat ${\cal H}^{{\mathrm dp}}$ perturbatively.
Following the formalism in the literature \cite{VANV48} 
for the $S$$=$$10$ system, 
we calculate to first order in ${\cal H}^{{\mathrm dp}}$ 
the root-mean-square deviation of the
resonant field from the center value $H_{\mathrm res}$ 
in Eq.~(\ref{eq:res_field}), assuming that the line is 
symmetric about the center value.
We do not account for the exact geometry of the sample for
this calculation. We use a lattice sum, assuming 
that dipoles are distributed on a simple cubic lattice, and 
the magnetic anisotropy axis is not particularly aligned 
along any of the edges of the lattice. The specific angles
of the magnetic anisotropy axis with the lattice axes do
not appreciably change the strength of the dipolar interactions.
The dipolar interactions make the linewidths decrease 
as $M_s$ decreases because the stronger
resonant fields lead to a more polarized system.  
%For the same reason, we expect that the dipolar interaction increases
%when EPR frequency decreases or temperature increases.

In this study, we emphasize that the line broadening effect due to 
hyperfine fields does not appear to be a significant factor. 
Indeed, for Fe$_8$, there is no hyperfine splitting due to 
the metal ion, since the dominant Fe isotope, $^{56}$Fe, 
has no nuclear spin.

The overall features of the linewidth, as a function of $M_s$,
are determined by the competition among these three effects:  
$D$-strain; $g$-strain; and dipolar interactions.
We have varied $\sigma_D$, $\sigma_g$, and the effective distance between 
the neighboring dipoles, $d$, within experimentally acceptable ranges 
in order to fit the experimental data. 
%Those parameters are
%approximately independent of temperature and EPR frequency.

\subsection{Mn$_{12}$}

Since the molecular magnet ${\mathrm Mn_{12}}$ has tetragonal symmetry, 
the ground-state single-spin Hamiltonian is, to lowest order,
\begin{eqnarray}
{\cal H}_{0} &=& -D S_z^2 - C S_z^4 - g \mu_B H_z S_z 
\end{eqnarray}
with $D$$=$$0.55k_B$, $C$$=$$1.17 \times 10^{-3}k_B$, and $g$$=$$1.94$.
\cite{BARR97} We perform the same analysis as in the previous subsection for
the available ${\mathrm Mn_{12}}$ data ($f$$=$$169~{\mathrm GHz}$,
$T$$=$$25~{\mathrm K}$, the field along the easy axis). 
See Fig.~\ref{FWHM}(d).
The differences are as follows: the resonant field is modified to 
\begin{eqnarray*}
\! \! H_{\mathrm res} &\equiv& 
\frac{\hbar \omega \!- \!D(2M_s-1) \! + \! C(4M_s^3-6M_s^2+4M_s-1)}{g \mu_B};
\end{eqnarray*}
dipoles are assumed to be 
distributed on a bcc lattice; and the anisotropy axis is along the
longer side of the sample.
 
For Mn$_{12}$, the hyperfine effect must be weak, because there is no
resolved hyperfine splitting in the EPR spectra. We attribute this
absence of hyperfine splitting to exchange narrowing due to fast
electron spin exchange over the metal-ion framework. Otherwise one
would expect a linewidth of over 1000 G, since each $^{55}$Mn nucleus
($I$$=$$\frac{5}{2}$) is expected to exhibit a hyperfine splitting
of about 200 G.\cite{AABB} Thus 12 almost equivalent $^{55}$Mn nuclei, 
together with the acetate protons alone should lead to an overall
width of over 1000 G. Notice that the measured smallest width
($M_s$$=$1 in Fig.~\ref{FWHM}(d)) is about 1000 G, which increases to 
over 2000 G toward the transition $M_s$$=$10$\rightarrow$9. 
The hyperfine effect would not be expected to change significantly
with the electronic spin quantum number $M_s$. We thus consider the
hyperfine effect as a lumped parameter representing the residual
linewidth, independent of $M_s$.

\section{Results}

\begin{figure}
\begin{center}
%\leavevmode
\epsfxsize=8.3cm
\epsfysize=5.cm
%\epsfbox{FWHM_Fe8.eps}
\epsfbox{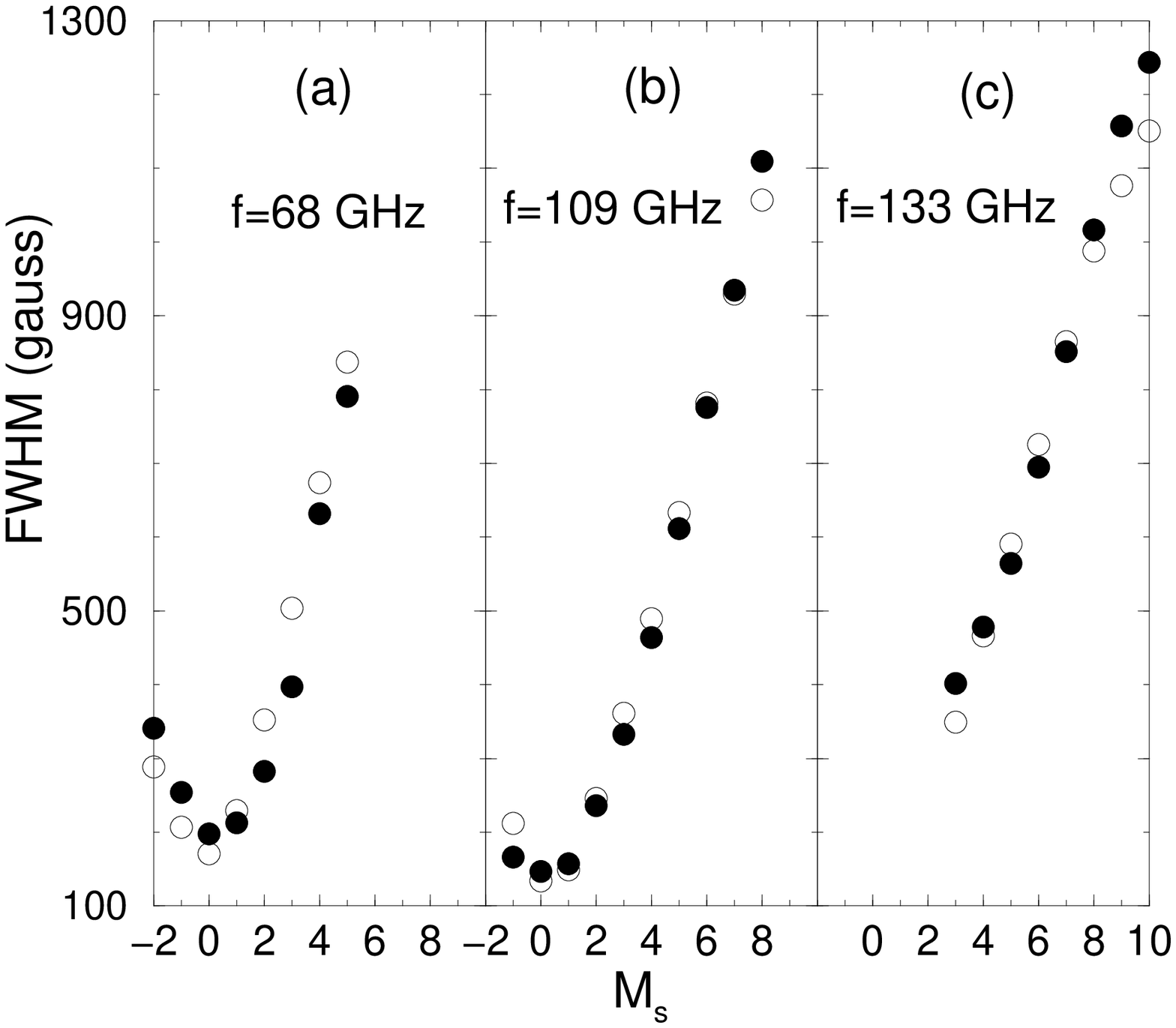}
\epsfxsize=8.3cm
\epsfysize=5.cm
\epsfbox{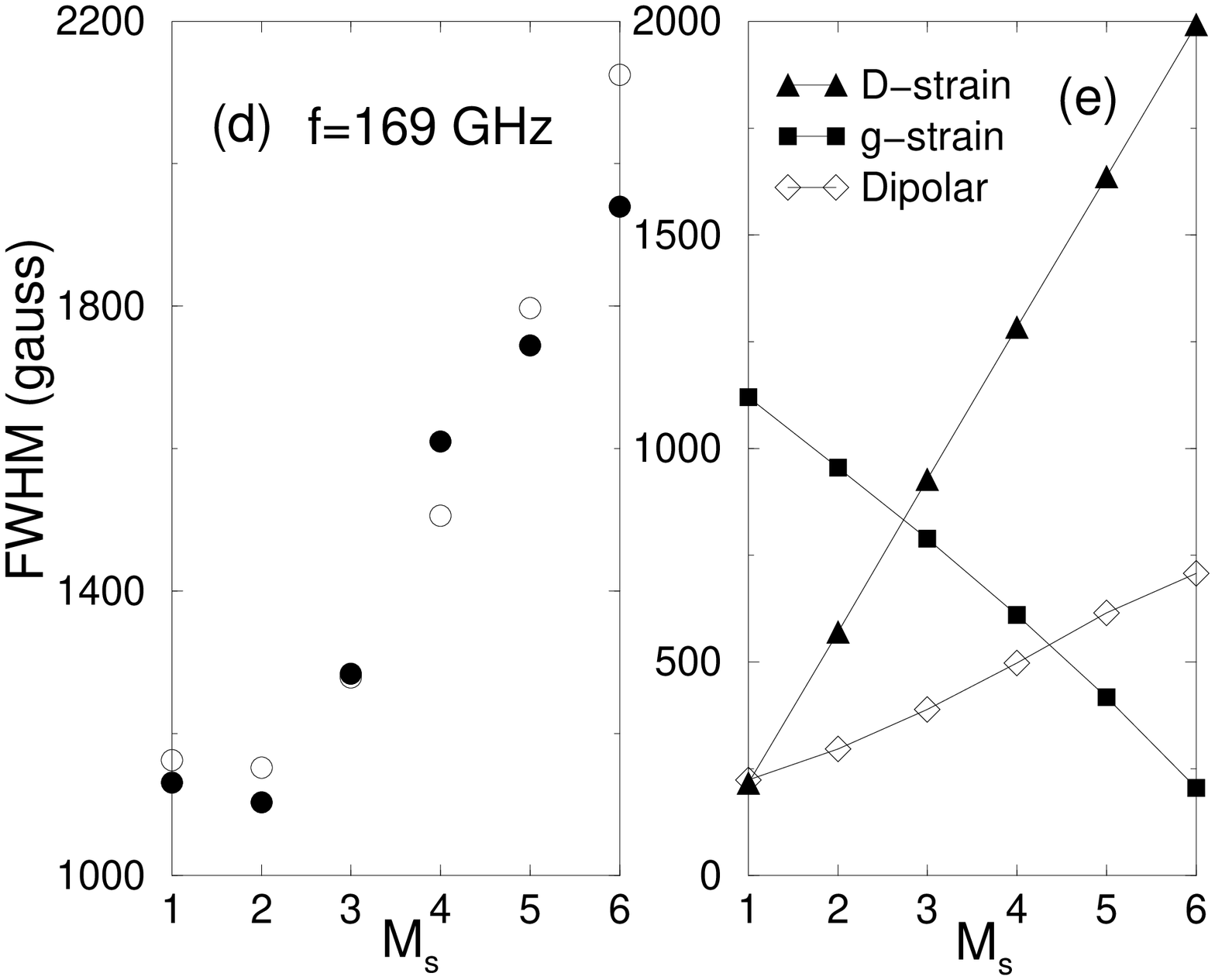}
\vspace{0.1in}
\caption{Experimental (filled circles) and theoretical (open circles)
Full Width at Half Maximum (FWHM) of the resonance
lines vs $M_s$ for (a),(b),(c) different frequencies at $T$$=$$10~{\mathrm K}$ 
for ${\mathrm Fe_8}$, and (d) at $T$$=$$25~{\mathrm K}$
for ${\mathrm Mn_{12}}$. (e) line broadening due to $D$-strain, $g$-strain,
and dipolar interactions vs $M_s$ for the Mn$_{12}$ sample in (d).
The standard deviations of $D$ and $g$, and the effective distance
between dipoles for (e) are given below.
Here $\vec{H} \, || \, \hat{z}$.
%The filled and open circles in (a)-(d) denote 
%the experimental and calculated FWHM, respectively.
For ${\mathrm Fe_8}$ (${\mathrm Mn_{12}}$), the fit is best 
when the standard deviation of $D$ is $0.01D$ ($0.02D$ and
the standard deviation of $g$ is $0.008g$) and the effective distance 
between dipoles is $12~{\mathrm \AA}$ ($14~{\mathrm \AA}$).}
\label{FWHM}
\end{center}
\end{figure}

For the ${\mathrm Fe_8}$ sample examined, the calculated linewidths
agree well with the experimental data at the measured frequencies
($f=68$, 89, 109, 113, 133, and 141~{GHz}) at $T$$=$$10~{\mathrm K}$,  
using $\sigma_D$$\approx$$0.01D$ and $d$$\approx$$12~{\mathrm \AA}$. 
Within acceptable ranges of $\sigma_D$, $\sigma_g$, and $d$,
no other combination of parameter values is able to
produce the same quality of fit simultaneously to all of the frequencies 
studied, although other parameter values may produce good fits for
a particular frequency.
Figures ~\ref{FWHM}(a), (b), and~(c) show the experimental and 
theoretical linewidths for ${\mathrm Fe_8}$ at three frequencies.
Our calculations show that the $D$-strain and the dipolar interactions 
are equally important for the linewidths of the sample, while 
the $g$-strain does not contribute significantly. 
%Because the distribution in $D$
%only results in {\it linear} $M_s$ dependence of the linewidths,
%a previously speculated $M_s^2$ dependence
%of the EPR linewidths for ${\mathrm Fe_8}$ and ${\mathrm Fe_4}$,
%\cite{BARR00} attributed to a spread in $D$ only, does not 
%adequately account for the measured experimental data 
%(Figs.~\ref{FWHM}(a),(b), and (c)). 
It was previously speculated that $M_s^2$ dependence of 
the EPR linewidths for ${\mathrm Fe_8}$ and ${\mathrm Fe_4}$ 
was due to the $D$-strain only.\cite{BARR00} However, this 
%A previously speculated $M_s^2$ dependence of the EPR linewidths 
%for ${\mathrm Fe_8}$ and ${\mathrm Fe_4}$,\cite{BARR00} 
cannot be explained by the $D$-strain alone, because the distribution in $D$
only results in a {\it linear} $M_s$ dependence of the linewidths.
The possibility of the presence of dipolar fields was mentioned
in a zero-field millimeter-wave experiment 
for ${\mathrm Fe_8}$.\cite{MUKH01}
The weak asymmetry of the linewidths about $M_s$$=$1 and $M_s$$=$0, 
shown in Figs.~\ref{FWHM}(a) and~(b), can be understood by combining
the $D$-strain effect with the dipolar interactions. The $D$-strain
produces linewidths symmetric about $M_s$$=$1 and $M_s$$=$0, while the dipolar
interactions yield narrower linewidths with smaller $M_s$ transitions,
which are observed at higher resonant fields. 
For a given $M_s$ transition, the linewidths decrease as the
frequency increases because of the dipolar interactions. 
Compare Fig.~\ref{FWHM}(a) with \ref{FWHM}(c).
The distribution in $D$ does not give frequency-
or resonant-field dependent linewidths, but the dipolar interactions do. 
This is another reason that the distribution in $D$ alone cannot explain
the measured linewidths.
%For example, Full Width at Half Maximum (FWHM) for $M_s$$=$$5$
%at $f$$=$$68~{\mathrm GHz}$ is about $800~{\mathrm gauss}$ while that at
%$f$$=$$133~{\mathrm GHz}$ is about $600~{\mathrm gauss}$.
We also considered distributions for other higher-order crystal-field 
anisotropy parameters ($E$ or fourth-order coefficients) and found that
they do not contribute substantially to the linewidths.
%although a distribution of the effective transverse anisotropy 
%plays an essential role to explain MQT in the recent tunneling model 
%\cite{CHUD01}.

For the Mn$_{12}$ sample examined, 
the calculated linewidths [Fig.~\ref{FWHM}(d)] 
are in a good agreement with the experimental
data for $\sigma_D$$\approx$$0.02D$, $\sigma_g$$\approx$$0.008g$, 
and $d$$\approx$$14~{\mathrm \AA}$. Figure~\ref{FWHM}(e) shows
separately the line broadening due to the $D$-strain, $g$-strain, and
dipolar interactions with the chosen standard deviations and
the effective distance between dipoles. 
For the Mn$_{12}$ sample, the distribution of $D$ is wider 
than that for the ${\mathrm Fe_8}$ sample. The $D$-strain effect 
is still dominant, but a $g$-strain effect must be included as well
to understand the linewidths for smaller $M_s$ transitions.
As shown in Fig.~\ref{FWHM}(e), the linewidths for 
$M_s$$=$2$\rightarrow$1 and $M_s$$=$1$\rightarrow$0 of
over 1000 G cannot be explained by the $D$-strain and the dipolar
interactions alone. It is essential to invoke the $g$-strain
effect. The dipolar interactions do not play as 
significant a role as for the ${\mathrm Fe_8}$ sample. 

\section{Conclusions}

We have presented our measured EPR spectra for the single-crystal
molecular magnets Fe$_8$ and Mn$_{12}$ as functions of energy eigenstate 
$M_s$ and EPR frequency when the field is along the easy axis. 
We also have calculated theoretically the EPR linewidths to 
quantitatively compare with the experimental data, using density-matrix 
equations along with Gaussian distributions in $D$ and $g$ and the 
dipolar interactions. Our calculated linewidths agree well with the
experimental data. Our analysis shows that for the examined Fe$_8$ 
sample the distribution
in $D$ and the dipolar interactions play key roles in explaining
the linewidths, while for the Mn$_{12}$ sample the distributions
in $D$ and $g$ are more important than the dipolar interactions.
From these, we draw the conclusion that the $D$-strain is universally
important for the EPR linewidths in molecular magnets, although
%In conclusion, our theoretical and experimental study of the dependence of 
%the EPR linewidths on $M_s$ show that a distribution in 
%%the uniaxial anisotropy parameter 
%$D$ seems to be generic in the molecular
%magnets ${\mathrm Fe_8}$ and ${\mathrm Mn_{12}}$, although the 
%%standard deviation 
%$\sigma_D$ varies between samples. 
other effects, such as $g$-strain and dipolar
interactions, can also contribute, depending on the material.
These results imply that the assumption made in the tunneling
model \cite{CHUD01} may be correct.
A distribution in $D$ may also facilitate multi-photon transitions 
through virtual states which are necessary to implement a quantum algorithm
for the molecular magnets ${\mathrm Mn_{12}}$ and ${\mathrm Fe_8}$.
\cite{LEUE00-3} Additionally we speculate that the $D$-strain effect 
may contribute to the widths of resonant steps in magnetization hysteresis 
loops. A more detailed report,\cite{SING} including the temperature
dependence of the linewidths at fixed frequency for both molecular magnets, 
will be published elsewhere.\cite{KPAR01B} \\

\begin{center}
{\textbf{Acknowledgments}}
\end{center}  
This work was partially funded by NSF Grant Nos.~DMR-9871455
and DMR-0103290, by Florida State University through CSIT and
MARTECH, and by Research Corporation (S.H.).

% tables follow here
%
% Here is an example of the general form of a table:
% Fill in the caption in the braces of the \caption{} command. Put the label
% that you will use with \ref{} command in the braces of the \label{} command.
% Insert the column specifiers (l, r, c, d, etc.) in the empty braces of the
% \begin{tabular}{} command.
%
% \begin{table}
% \caption{}
% \label{}
% \begin{tabular}{}
% \end{tabular}
% \end{table}

%\end{multicols}


\begin{references}
\bibitem{SING}
%It will include a quantitative study 
%of ${\mathrm Mn_{12}}$ data for several frequencies and the temperature
%dependence of the linewidths at fixed frequency for both molecular magnets.
Extension of this model to a full line-shape analysis
indicates that it might be possible to distinguish
between the distributions due to various types of physical
mechanisms, including edge dislocations.
\end{references}
\end{document}